\documentclass{Interspeech}
\usepackage{tcolorbox}
\usepackage{multirow}

\interspeechcameraready

\title{Speech LLMs in Low-Resource Scenarios: Data Volume Requirements and the Impact of Pretraining on High-Resource Languages}

\author[affiliation={1,2}]{Seraphina}{Fong}
\author[affiliation={2}]{Marco}{Matassoni}
\author[affiliation={2}]{Alessio}{Brutti}

\affiliation{Department of Information Engineering and Computer Science}{University of Trento}{Italy}
\affiliation{Center for Augmented Intelligence}{Fondazione Bruno Kessler}{Italy}

\email{meiyueseraphina.fong@unitn.it, matasso@fbk.eu, brutti@fbk.eu}

\usepackage{comment}

\begin{document}

\maketitle


\begin{abstract}
Large language models (LLMs) have demonstrated potential in handling spoken inputs for high-resource languages, reaching state-of-the-art performance in various tasks. However, their applicability is still less explored in low-resource settings. This work investigates the use of Speech LLMs for low-resource Automatic Speech Recognition using the SLAM-ASR framework, where a trainable lightweight projector connects a speech encoder and a LLM. Firstly, we assess training data volume requirements to match Whisper-only performance, re-emphasizing the challenges of limited data. Secondly, we show that leveraging mono- or multilingual projectors pretrained on high-resource languages reduces the impact of data scarcity, especially with small training sets. Using multilingual LLMs (EuroLLM, Salamandra) with whisper-large-v3-turbo, we evaluate performance on several public benchmarks, providing insights for future research on optimizing Speech LLMs for low-resource languages and multilinguality.
\end{abstract}

\section{Introduction}
\label{section:intro}
Low-resource settings, characterized by limited data or lack of manual annotations, remain a major challenge for speech technologies in production environments \cite{LamYeeMui2023MultilingualMW}. For example, current Automatic Speech Recognition (ASR) models are limited by the amount and diversity of training data required, supporting only about 100 of the more than 7,000 languages spoken globally \cite{fatehi2024overview,liu2024exploration,chiang2023whisperhakka}. As a result, the scarcity of high-quality training data continues to hinder advances in speech and language technologies for low-resource scenarios \cite{liu2024exploration,chiang2023whisperhakka,roger2022deep,Cheng2024ExploringTI}. Therefore, the need for technologies that support underrepresented languages is increasingly important, to not only preserve linguistic and cultural heritage, but to also ensure that a broader range of users worldwide have access to modern technologies.

Originally developed for text generation, large language models (LLM) have demonstrated increasing versatility in performing an expanding range of tasks \cite{liu2024visual,wu2023multimodal,bai2024seed,fathullah2024prompting}. Of particular interest to the present work is how they have been extended to process speech and audio, enabling new possibilities for speech-related tasks. Recent LLM-based ASR approaches have implemented a trainable projector to integrate speech encoders with LLMs to improve ASR performance by leveraging the extensive knowledge of LLMs  \cite{mittal2024salsa,mundnich2024zero,tang2024salmonn,chu2023qwen,wu2023decoder}. 
The projector's role is to align speech embeddings from the speech foundation encoder with the LLM's embedding space, enabling an LLM to handle speech input alongside textual prompts. 
The SLAM-ASR \cite{ma2024slamasr} approach demonstrated that implementing a lightweight layer as the projector between off-the-shelf speech foundation encoders and LLMs could achieve state-of-the-art results on the English Librispeech dataset \cite{panayotov2015librispeech}. However, most emerging research in this domain predominantly focuses on high-resource languages (e.g., English, Mandarin Chinese), and therefore have large amounts of data available for training (e.g., 960 hours of Librispeech for SLAM-ASR \cite{ma2024slamasr,kumar2024performance}, 1000 hours of data per language in \cite{wu2023decoder}). Other studies have investigated the impact of factors such as data characteristics or computational resources~\cite{kumar2024performance}. In particular, \cite{cappellazzo2024largelanguagemodelsstrong} demonstrates the crucial role of training data volumes for LLM-based audio-visual ASR.
However, it remains unclear how well the LLM-based framework performs for low-resource settings, as data scarcity can impact all components of the pipeline. For example, the speech encoder may learn less robust representations, the LLM may generate less accurate transcripts due to limited exposure to the target language, and the projector may struggle to effectively align the different modality embeddings. Furthermore, in low-resource settings, there is often little to no data available not only for ASR, but also for other speech-related tasks such as spoken language understanding and spoken question answering. Given these challenges, it is essential to investigate the viability of Speech LLMs in low-resource scenarios. In this work, we use ASR as a case study. Therefore, our research questions (RQ) are as follows:

\begin{tcolorbox}[colback=green!5]
\textbf{(}$\mathbf{RQ1}$\textbf{)} \textit{How many hours of training data are needed to effectively train a linear projector?} \\
\textbf{(}$\mathbf{RQ2}$\textbf{)} \textit{Can linear projectors pretrained on a high-resource language be fine-tuned on another language to mitigate the impact of data scarcity in low-resource ASR?}
\end{tcolorbox}

To address these questions, we progressively increase the amount of data used to train a linear projector within the SLAM-ASR framework, starting from 10 to 252 hours of the Common Voice (CV) Italian dataset \cite{commonvoice:2020}, to determine the amount of data needed to match or exceed the performance of using a Whisper-only framework \cite{radford2023robust}. We subsequently explore the effects of leveraging a projector pretrained on English data (Librispeech 100 \cite{panayotov2015librispeech} and 100 hours of CV English) or 200 hours of CV Spanish data, fine-tuning them on varying amounts of Common Voice Italian data (10, 15, 100, and 200 hours) to assess its potential adaptability and performance in low-resource settings. We further extend this analysis to a real-world low-to-medium resource scenario with 10-15 hours of Galician data, where we also bootstrap a multilingual projector pretrained with a combination of English, Spanish, and Italian data. Consequently, we also investigate the impact of the selected LLM using two open-source multilingual LLMs (EuroLLM 1.7B~\cite{Martins2024EuroLLMML} and Salamandra 2B~\cite{gonzalezagirre2025salamandratechnicalreport}), as well as reinforce previous findings \cite{kumar2024performance} of SLAM-ASR performance in cross-domain evaluation settings. 

Overall, this work presents the following contributions to provide insights into the applicability of Speech LLMs in low-resource scenarios, using SLAM-ASR as a case-study framework:
\begin{itemize}
    \item Achieving performance comparable to Whisper-only models requires 200 hours of training data, highlighting the persistent challenge of data scarcity in low-resource settings.
    \item Fine-tuning projectors pretrained on high-resource languages enhances cross-domain performance, especially when the fine-tuning dataset is limited (10–15 hours). Performance is further improved when leveraging projectors pretrained on multiple languages (i.e., a ``multilingual" projector).
    \item The choice of multilingual LLM impacts performance, with EuroLLM 1.7B performing better than Salamandra 2B.
\end{itemize}

\section{Methodology}

\begin{figure}[ht!]
  \centering
  \includegraphics[width=0.85\columnwidth]{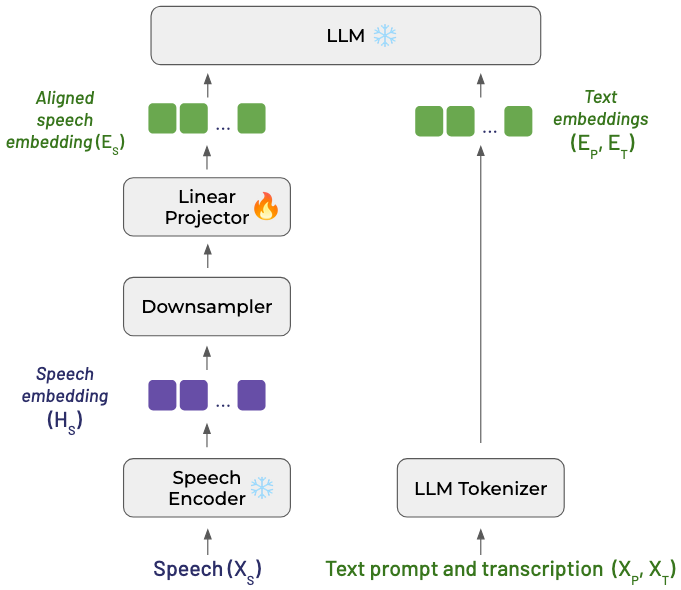}
  \caption{The SLAM-ASR framework \cite{ma2024slamasr} incorporates a trainable lightweight projector to align speech features \(H_S\) from a frozen speech encoder within the embedding space of a frozen LLM. The projector's output features \(E_S\) are combined with the corresponding embeddings of a text prompt \(E_P\) (and, during training, of a transcription \(E_T\)) to generate the predicted transcription of the input speech signal.}
  \label{fig:slam_asr_framework}
\end{figure}

We employ the popular SLAM-ASR framework and code-base for our experimental analysis~\cite{ma2024slamasr}, chosen for its open-source accessibility and strong performance on English benchmarks. As depicted in Figure \ref{fig:slam_asr_framework}, SLAM-ASR consists of a frozen speech foundation encoder, a trainable linear projector, and a frozen LLM. The input speech signal, its corresponding transcript, and the text prompt are denoted as $X_s$,  \(X_T\), and \(X_P\), respectively. For each input sample \(X_s\), speech embeddings \(H_s\) are first extracted from the speech signal \(X_s\) via the speech encoder where:

\begin{equation}
  H_s = Encoder(X_s)
  \label{equation:speech_encoder_eq}
\end{equation}

\noindent We consider the speech encoder from the \textit{Whisper-large-v3-turbo} model\footnote{\url{https://huggingface.co/openai/whisper-large-v3-turbo}}\cite{radford2023robust}, a state-of-the-art transformer-based system for multilingual speech recognition.  The extracted speech embeddings are downsampled by a factor of \textit{k} = 5 to address the length discrepancy between speech and text features. A linear projector then transforms the downsampled \(H_s\) into \(E_s\) to align with the LLM’s input embedding dimension. Following the original SLAM-ASR implementation, the linear projector consists of a single hidden layer followed by ReLU activation and a regression layer, denoted as:

\begin{equation}
  E_s = Linear(ReLU(Linear(Downsample(H_s)))
  \label{equation:linear_projector_eq}
\end{equation}

The transcript \(X_T\) and prompt \(X_P\) are tokenized by the LLM's tokenizer to obtain the corresponding transcript embedding \(E_T\) and prompt embedding \(E_P\). During both training and decoding, the embeddings are concatenated to form the final input embedding. Specifically, during training, \(E_S\), \(E_P\), and \(E_T\) are concatenated, while during decoding, only \(E_S\) and \(E_P\) are concatenated before being passed into the LLM to generate the predicted output transcript. We consider two open-source LLMs that are multilingual on European languages (EuroLLM 1.7B\footnote{\url{https://huggingface.co/utter-project/EuroLLM-1.7B}} and Salamandra 2B\footnote{\url{https://huggingface.co/BSC-LT/salamandra-2b}}). 

\begin{figure}[ht!]
  \centering
  \includegraphics[width=0.99\columnwidth]{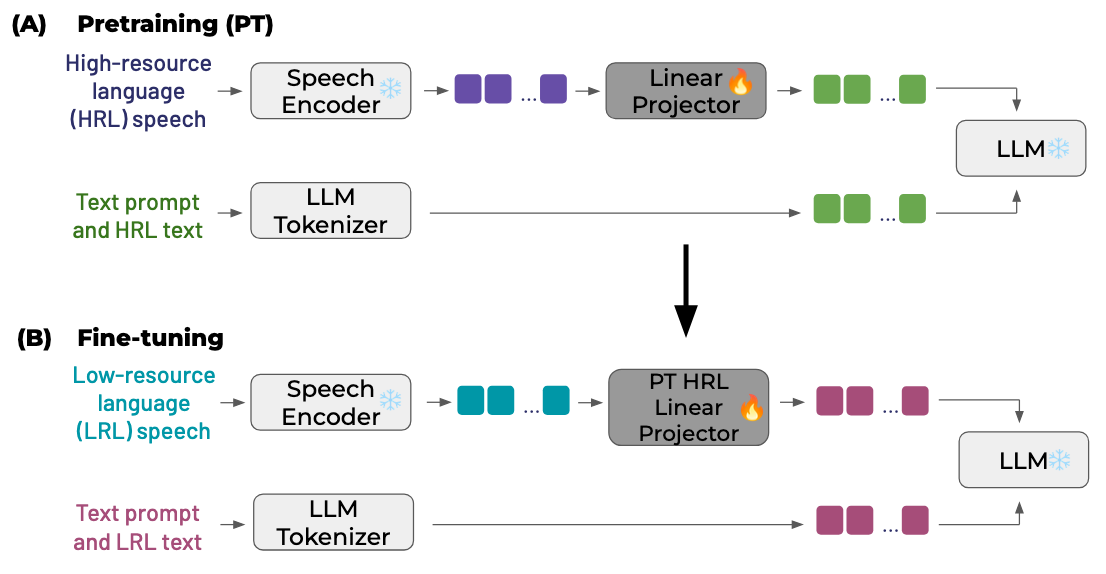}
  \caption{The pretraining and finetuning framework. A lightweight projector is first pretrained (PT) on a high-resource language (HRL; Panel A) and then finetuned on low-resource language data (LRL; Panel B).}
  \label{fig:pretrain_finetune_framework}
\end{figure}

We first use Italian, a well-resourced language, to simulate low-resource settings by artificially limiting the amount of data for training. This choice allows us to disentangle possible issues related to the quality of the LLM and the speech encoder. We then examine the potential of pretraining a projector on high-resource language data (Figure \ref{fig:pretrain_finetune_framework}A), and then fine-tuning them to evaluate its applicability in low-resource settings (Figure \ref{fig:pretrain_finetune_framework}B).

\section{Experimental Setup}

\subsection{Datasets}

Our experiments are conducted using the Librispeech (LS) \cite{panayotov2015librispeech}, Common Voice 20.0 (CV) \cite{commonvoice:2020}, and Fleurs (FL) datasets \cite{FLEURS}. For LS, we use the \textit{train-clean-100 hours}, \textit{dev-other}, and \textit{test-clean} subsets. For CV and FL, we use the English (EN), Italian (IT), Spanish (ES\_419), and Galician (GL) data. With the Common Voice datasets, we create training data subsets totaling between 10 and 252 hours of data for IT, 100 hours for EN, 200 hours for ES, and 10 to 15 hours for GL, randomly selecting signals less than 20 seconds (14 for GL) due to computational limitations. 

\subsection{Training Configuration}

All models were trained using the code-base provided by the official SLAM-ASR Github repository\footnote{\url{https://github.com/X-LANCE/SLAM-LLM/tree/main/examples/asr_librispeech}}. For all experiments, we utilize \textit{Whisper-large-v3-turbo} as the speech encoder and a linear projector. We use two multilingual LLMs (EuroLLM 1.7B and Salamandra 2B) to compare the resulting performance as the choice of LLM has been found to impact it \cite{kumar2024performance}. 

Our training setup follows the original SLAM-ASR implementation \cite{ma2024slamasr} in several aspects. We use the AdamW optimizer (learning rate = 1e-4) and no weight decay for the trainable parameters. A learning rate scheduler with a 1,000-step warmup followed by a fixed maximum learning rate throughout training is applied. Training is set for a maximum of 100,000 training steps, which is stopped early if the validation loss does not decrease. We also use beam search decoding with \textit{beam size} = 4. In a few experiments, we finetune the query and value projection layers of EuroLLM 1.7B with low-rank matrices using the repository's default parameters (\textit{r} = 8, $\alpha$ = 32, drop-out = 0.05) while training the projector. This added 1.38M tunable parameters alongside the 17.31M parameter linear projector.

Due to the availability of computational resources, we differ in some areas: training is conducted on 100 hours of LS for three epochs and on other language datasets for six epochs. The batch size was set to four. All models were trained with one NVIDIA Ada Lovelace L40S GPU. Additionally, the context length of the model was adjusted accordingly based on the LLM used (i.e., 4096 for EuroLLM 1.7B and 8192 for Salamandra 2B).

For our monolingual projectors, we provided a language-specific prompt to the LLM, e.g., \textit{``Transcribe [LANGUAGE] speech to text"}. For instance, the prompt used for Italian data was \textit{``Transcribe Italian speech to text"}. Preliminary experiments showed that a language-specific prompt, compared to the general one utilized in the original SLAM-ASR implementation (\textit{``Transcribe speech to text"}), affected the model's performance. For instance, with English, the language-specific prompt reduced the WER on LS Test Clean from 5.9\% to 4.6\%.

\section{Results}
This section is organized according to the research questions posed in Section \ref{section:intro}. All models are evaluated using the Word Error Rate (WER) metric to measure the similarity between the ground-truth transcripts versus a given model's predicted transcriptions. 

\subsection{RQ1: Training Data Requirements for Effective Linear Projector Training}

Table \ref{tab:results_italian_training_material_amount} includes the results of the linear projector's performance as the amount of training data from the CV IT dataset increases from 10 hours to 252 hours. The results demonstrate that increasing the quantity of training data consistently improves the overall performance of the model, regardless of the LLM used within the SLAM-ASR framework. However, using EuroLLM 1.7B consistently outperformed Salamandra 2B, highlighting the significant impact of LLM selection on downstream ASR performance. Note that the performance gap between Salamandra and EuroLLM tends to close as more data are available. This could be related to the larger context required by Salamandra. When comparing the performance of the SLAM-ASR framework to standalone whisper-only models, the configuration with EuroLLM 1.7B and 200 or 252 hours of training data obtains a WER of 6.4\% and 6.1\% respectively on CV IT, outperforming \textit{Whisper-large-v3-turbo} (7.1\%). This suggests that the SLAM-ASR framework can provide slightly better performance than whisper-only models, but requires a significant amount of training data to achieve these gains, in spite of the extremely lightweight projector being trained. Moreover, it does not outperform a \textit{Whisper-large}- or \textit{Whisper-large-v3-turbo}-only set-up on Fleurs IT (WER = 4.7\% and 5.8\%, respectively).

Performance is notably better on in-domain (CV IT) data compared to out-of-domain (FL IT) data across all configurations. For example, with 200 hours of CV IT training data with EuroLLM 1.7B, the WER on CV IT is 6.4\% but 13.2\% on FL IT. This gap demonstrates the challenge that the SLAM-ASR framework has when generalizing across domains, a finding aligned with existing research~\cite{kumar2024performance}.

Following \cite{kumar2024performance}'s suggestion that LoRA fine-tuning of the LLM could improve the alignment between speech and text tokens, additional experiments with 15 and 100 hours of training data were conducted. These experiments were performed solely with EuroLLM 1.7B, given its better performance over Salamandra 2B. Our findings confirmed that LoRA can be beneficial. However, we observed only a marginal difference in in-domain performance: a reduction of 0.3\% WER with 15 hours of data and 0.5\% with 100 hours of data. Note, however, that LoRA improves out-of-domain performance on FL IT, reducing WER by 1.2\% (with 15 hours of data) and 2.5\% (with 100 hours of data). Given the marginal changes in performance, we do not employ LoRA in subsequent experiments.

\begin{table}[th]
  \caption{Results obtained over the CommonVoice (CV) Italian (IT) test set and the Fleurs IT test set. We progressively increase the amount of data used to train a linear projector within the SLAM-ASR framework, starting from 10 to 252 hours of the CV IT training set.
  Bolded values with an asterisk (*) indicate results surpassing a setup consisting only of \textit{Whisper-large-v3-turbo} (CV IT WER = 7.1\%). For reference, \textit{Whisper-large} has a CV IT WER = 6.5\% and a Fleurs IT WER = 4.7\%, while \textit{Whisper-large-v3-turbo} has a Fleurs IT WER = 5.8\%.}
  \label{tab:results_italian_training_material_amount}
  \centering
  \resizebox{\columnwidth}{!}{
  \begin{tabular}{l r c c c c }
    \toprule
      & &\multicolumn{2}{c}{\textbf{EuroLLM 1.7B}}&\multicolumn{2}{c}{\textbf{Salamandra 2B}}\\
     \textbf{Train}& \textbf{hours}&\textbf{CV IT}&\textbf{Fleurs IT}& \textbf{CV IT}&\textbf{Fleurs IT}\\
    \midrule
    Fleurs & 9 & -&18.0& -&22.5\\
    \hline
    CV &10 & 14.0&35.2& 33.6&111.4\\
    CV &15 & 13.9& 34.0 & 26.0& 146.3\\
  \multicolumn{2}{c}{+LoRA} & 13.6& 32.8 & - & -\\
     CV &20& 11.4& 31.6& 19.3& 68.3\\
      CV &50& 9.2& 20.6& 12.2& 56.6\\
     CV &100& 7.6& 16.3& 10.4& 80.6\\
     \multicolumn{2}{c}{+LoRA} & 7.1 & 13.8 & - &- \\
     CV &200& \textbf{6.4*}& 13.2& 7.7& 65.7\\
     CV &252& \textbf{6.1*}& 12.3& - & - \\
    \bottomrule
  \end{tabular}}
\end{table}

\subsection{RQ2: Bootstrapping pretrained linear projectors towards low-resource 
languages}
\label{subsection:rq2-bootstrapping}

\begin{table*}[th!]
  \caption{Results on Common Voice Italian (CV IT)  and Fleurs (FL) IT test sets after finetuning a projector pretrained on English LibriSpeech 100 (LS100 EN$\xrightarrow{}$), 100 hours of Common Voice (CV) English data (CV100 EN$\xrightarrow{}$), or 200 hours of Common Voice Spanish data (CV200 ES$\xrightarrow{}$) using 10, 15, 100, or 200 hours of CV IT training data. Performance is compared to a projector trained from scratch on CV IT (Scratch).}
  \label{tab:results_finetuning}
  \centering
  \begin{tabular}{l c c c c cccc}
    \toprule
     &\multicolumn{2}{c}{\textbf{Scratch}}&\multicolumn{2}{c}{\textbf{LS100 EN$\xrightarrow{}$}}&\multicolumn{2}{c}{\textbf{CV100 EN$\xrightarrow{}$}}&\multicolumn{2}{c}{\textbf{CV200 ES$\xrightarrow{}$}}\\
      \textbf{Training data (hours)}& \textbf{CV IT}&\textbf{FL IT}& \textbf{CV IT}&\textbf{FL IT} & \textbf{CV IT}&\textbf{FL IT}& \textbf{CV IT}&\textbf{FL IT}\\
    \midrule
     CV (10)& 14.0 & 35.2& 12.0& 16.1 & 9.8 & 17.1& 8.6& 20.3\\
     CV (15)& 13.9 & 34.0& 12.8& 15.7 & 10.0 & 16.9 & 8.8 & 16.1\\
     CV (100)& 7.6& 16.3& 8.1& 12.6& 7.7& 16.4& 7.3 & 17.6\\
     CV (200)& 6.4& 13.2& 6.6 & 11.6 & 6.7 & 15.8 & 6.4 & 17.3\\
     \hline
  \end{tabular}  
\end{table*}

\begin{table}[th!]
  \caption{Results on the Common Voice (CV) and Fleurs (FL) Galician (GL) test sets after finetuning a projector pretrained on 200 hours of CV Spanish (CV200 ES$\xrightarrow{}$), Italian (CV200 IT$\xrightarrow{}$), or a combination of the two former datasets with Librispeech 100 (MULTI$\xrightarrow{}$), using 10–15 hours (h) of CV GL training data. Performance is compared to a projector trained from scratch on CV GL (Scratch). For reference, standalone Whisper-large-v3-turbo has a CV GL WER = 16.2\% and FL GL WER = 14.8\%.}
  \label{tab:results_finetuning_galician}
  \centering
  \resizebox{\columnwidth}{!}{%
  \begin{tabular}{lllllllll}
  \toprule
     & \multicolumn{2}{c}{\textbf{Scratch}}& \multicolumn{2}{c}{\textbf{CV200 ES$\xrightarrow{}$}} & \multicolumn{2}{c}{\textbf{CV200 IT$\xrightarrow{}$}} & \multicolumn{2}{c}{\textbf{MULTI$\xrightarrow{}$}}\\
      \textbf{Train(h)}&  \textbf{CV}& \textbf{FL}& \textbf{CV}& \textbf{FL} & \textbf{CV}&\textbf{FL} & \textbf{CV}&\textbf{FL}\\
    \midrule
     CV (10)&  18.6& 43.0& 13.9& 27.1 & 15.1 &26.8 & 13.3&19.4\\
     CV (15)&  17.5& 42.4& 12.4& 22.0 & 13.5& 22.1  & 12.4 & 15.8\\
     \hline
  \end{tabular}
  }  
\end{table}

Results in Table~\ref{tab:results_italian_training_material_amount} show that at least 100-200 hours of labeled data are needed to achieve state-of-the-art ASR performance. This amount may be prohibitive in several languages. Therefore, we investigate a transfer learning approach by pretraining the projector on a different language. Given its superior performance, we only consider EuroLLM 1.7B in the next experiments. Table \ref{tab:results_finetuning} compares performance across various bootstrapping setups using models pretrained on high-resource datasets (100–200 hours) and evaluated on Italian datasets.  Models pretrained on English (LS100 EN$\xrightarrow{}$, CV100 EN$\xrightarrow{}$) and Spanish (CV200 ES$\xrightarrow{}$) fairly consistently outperformed the baseline ``Scratch" model trained solely on CV Italian.

In low-resource settings, such as when only 10 or 15 hours data are available, leveraging pretrained models appears to provide substantial gains both within in- and out-of-domain contexts. When fine-tuned on smaller datasets (e.g., CV IT with 10 or 15 hours of training data), the pretrained projectors demonstrate a significant improvement in WER over a model trained from scatch. For example, on CV IT with 10 hours of training data, the WER drops from 14.0\% (Scratch) to 9.8\% (CV100 EN$\xrightarrow{}$) and 8.6\% (CV200 ES$\xrightarrow{}$). Similarly, on FL IT with 10 hours of training data, the WER decreases from 35.2\% (Scratch) to 17.1\% (CV100 EN$\xrightarrow{}$) and 20.3\% (CV200 ES$\xrightarrow{}$). These findings suggest that pretraining may help generalize the model. However, as the amount of fine-tuning data increases to 100-200 hours, the advantage of fine-tuning a pretrained model diminishes with smaller performance gains. When fine-tuning using 200 hours of CV IT, the WER of the Scratch model is 6.4\%, which aligns with the performance of LS100 EN$\xrightarrow{}$ at 6.6\%. 

It is also worth noting that performance appears to be dependent on the language, semantic domain, and acoustic properties of the data used to pretrain the projector. In the case of CV IT, starting from a projector pretrained on CV gives better results. Moreover, using Spanish data is better than employing a projector trained on English as the acoustic similarity with Italian is higher for the former. Interestingly, we observe that bootstrapping the projector from an English pretrained model (LS100/CV100 EN$\xrightarrow{}$) has more performance gains in the out-of-domain context compared to Spanish.  
Future work could consider further exploring linguistic factors, multilingual pretraining, or other methods to enhance cross-lingual transfer, especially between closely related languages to further optimize outcomes.

\subsection{Galician: a low-resource case study}
Building on the findings in subsection \ref{subsection:rq2-bootstrapping}, we extend our bootstrapping analysis to Galician (GL) to assess the applicability of bootstrapping with an actual low resource language. Table \ref{tab:results_finetuning_galician} includes the results of fine-tuning a projector pretrained either on 200 hours of Common Voice Spanish (CV200 ES$\xrightarrow{}$) or Italian data (CV200 IT$\xrightarrow{}$), or on multiple language datasets (MULTI$\xrightarrow{}$; LibriSpeech 100, CV200 ES, and CV200 IT), compared to a projector trained from scratch on 10 or 15 hours of CV GL (Scratch). These findings align with our analysis in subsection \ref{subsection:rq2-bootstrapping}, demonstrating that transfer learning from a related high-resource language enhances performance over training from scratch, even with limited Galician data. Moreover, we find that a multilingual projector (MULTI$\xrightarrow{}$) can further improve performance and generalization capabilities. With only 10 hours of GL finetuning data, a fine-tuned multilingual projector achieves a WER of 13.3\% and 19.4\% on CV GL and FL GL, respectively, versus 18.6\% and 43\% when trained solely from Scratch. While benefits are still observed at 15 hours of data (WER 12.4\% CV GL and 15.8\% FL GL), they are less pronounced.

\section{Conclusion}

This study aimed to investigate the potential of SLAM-ASR, a recent Speech LLM-based framework, in low-resource scenarios by assessing the data volume requirements and the impact of high-resource pretraining and fine-tuning on data scarcity. Using \textit{Whisper-large-v3-turbo} as the speech encoder, a linear projector, and a multilingual LLM, results indicate that at least 100-200 hours of training data are needed for the SLAM-ASR framework to match the performance of Whisper-only models, reinforcing the difficulty of training in low-resource settings. Findings however showed that pretraining on high-resource languages can significantly improve cross-domain performance, especially with smaller fine-tuning datasets (10-15 hours).

Overall, Speech LLMs, in the context of the SLAM-ASR framework, show potential in low-resource scenarios through pretraining and fine-tuning strategies. Although its reliance on larger amounts of training data and challenges with cross-domain performance present areas for further development, these findings highlight avenues for future research towards adapting and optimizing the framework for low-resource languages.

\section{Acknowledgements}
This work has received funding from the European Union’s Horizon Europe research and innovation programme under the project ELOQUENCE (Grant Agreement No. 101135916).

\bibliographystyle{IEEEtran}
\bibliography{ref}

@inproceedings{radford2023robust,
author = {Radford, Alec and Kim, Jong Wook and Xu, Tao and Brockman, Greg and McLeavey, Christine and Sutskever, Ilya},
title = {Robust speech recognition via large-scale weak supervision},
year = {2023},
booktitle = {40th International Conference on Machine Learning},
articleno = {1182},
numpages = {27},
location = {Honolulu, Hawaii, USA},
series = {ICML'23}
}

@inproceedings{commonvoice:2020,
  author = {Ardila, R. and Branson, M. and Davis, K. and Henretty, M. and Kohler, M. and Meyer, J. and Morais, R. and Saunders, L. and Tyers, F. M. and Weber, G.},
  title = {Common Voice: A Massively-Multilingual Speech Corpus},
  booktitle = {Proceedings of the 12th Conference on Language Resources and Evaluation (LREC 2020)},
  pages = {4211--4215},
  year = 2020
}

@article{LamYeeMui2023MultilingualMW,
  title={{Multilingual Models with Language Embeddings for Low-resource Speech Recognition}},
  author={L{\'e}a-Marie Lam-Yee-Mui and Waad Ben Kheder and Viet Bac Le and Claude Barras and Jean-Luc Gauvain},
  journal={2nd Annual Meeting of the ELRA/ISCA SIG on Under-resourced Languages},
  year={2023},
  url={https://api.semanticscholar.org/CorpusID:264886709}
}

@INPROCEEDINGS{FLEURS,
  author={Conneau, Alexis and Ma, Min and Khanuja, Simran and Zhang, Yu and Axelrod, Vera and Dalmia, Siddharth and Riesa, Jason and Rivera, Clara and Bapna, Ankur},
  booktitle={IEEE Spoken Language Technology Workshop }, 
  title={{FLEURS: FEW-Shot Learning Evaluation of Universal Representations of Speech}}, 
  year={2023},
  volume={},
  number={},
  pages={798-805},
  
}

@article{liu2024exploration,
  title={{Exploration of Whisper fine-tuning strategies for low-resource ASR}},
  author={Liu, Yunpeng and Yang, Xukui and Qu, Dan},
  journal={EURASIP Journal on Audio, Speech, and Music Processing},
  
  number={1},
  pages={29},
  year={2024},
  publisher={Springer}
}

@inproceedings{chiang2023whisperhakka,
  title={{WhisperHakka: A Hybrid Architecture Speech Recognition System for Low-Resource Taiwanese Hakka}},
  author={Chiang, Ming-Hsiu and Lai, Chien-Hung and Chiu, Hsuan-Sheng},
  booktitle={35th Conference on Computational Linguistics and Speech Processing },
  pages={390--396},
  year={2023}
}

@article{roger2022deep,
  title={{Deep neural networks for automatic speech processing: a survey from large corpora to limited data}},
  author={Roger, Vincent and Farinas, J{\'e}r{\^o}me and Pinquier, Julien},
  journal={EURASIP Journal on Audio, Speech, and Music Processing},
  
  number={1},
  pages={19},
  year={2022},
  publisher={Springer}
}

@article{fatehi2024overview,
  title={{An overview of high-resource automatic speech recognition methods and their empirical evaluation in low-resource environments}},
  author={Fatehi, Kavan and Torres, Mercedes Torres and Kucukyilmaz, Ayse},
  journal={Speech Communication},
  pages={103--151},
  year={2024},
  publisher={Elsevier}
}

@inproceedings{panayotov2015librispeech,
  title={{Librispeech: an asr corpus based on public domain audio books}},
  author={Panayotov, Vassil and Chen, Guoguo and Povey, Daniel and Khudanpur, Sanjeev},
  booktitle={IEEE international conference on acoustics, speech and signal processing},
  pages={5206--5210},
  year={2015},
  organization={IEEE}
}

@inproceedings{tang2024salmonn,
  title={{SALMONN: Towards Generic Hearing Abilities for Large Language Models}},
  author={Tang, Changli and Yu, Wenyi and Sun, Guangzhi and Chen, Xianzhao and Tan, Tian and Li, Wei and Lu, Lu and Zejun, MA and Zhang, Chao},
  year={2024},
  booktitle={The Twelfth International Conference on Learning Representations}
}

@inproceedings{wu2023decoder,
  title={{On decoder-only architecture for speech-to-text and large language model integration}},
  author={Wu, Jian and Gaur, Yashesh and Chen, Zhuo and Zhou, Long and Zhu, Yimeng and Wang, Tianrui and Li, Jinyu and Liu, Shujie and Ren, Bo and Liu, Linquan and others},
  booktitle={IEEE Automatic Speech Recognition and Understanding Workshop},
  pages={1--8},
  year={2023},
  organization={IEEE}
}

@article{ma2024slamasr,
  title={{An Embarrassingly Simple Approach for LLM with Strong ASR Capacity}},
  author={Ma, Ziyang and Yang, Guanrou and Yang, Yifan and Gao, Zhifu and Wang, Jiaming and Du, Zhihao and Yu, Fan and Chen, Qian and Zheng, Siqi and Zhang, Shiliang and others},
  journal={arXiv preprint arXiv:2402.08846},
  year={2024}
}

@article{mundnich2024zero,
  title={{Zero-resource speech translation and recognition with LLMs}},
  author={Mundnich, Karel and Niu, Xing and Mathur, Prashant and Ronanki, Srikanth and Houston, Brady and Elluru, Veera Raghavendra and Das, Nilaksh and Hou, Zejiang and Huybrechts, Goeric and Bhatia, Anshu and others},
  journal={arXiv preprint arXiv:2412.18566},
  year={2024}
}

@article{chu2023qwen,
  title={{Qwen-audio: Advancing universal audio understanding via unified large-scale audio-language models}},
  author={Chu, Yunfei and Xu, Jin and Zhou, Xiaohuan and Yang, Qian and Zhang, Shiliang and Yan, Zhijie and Zhou, Chang and Zhou, Jingren},
  journal={arXiv preprint arXiv:2311.07919},
  year={2023}
}

@article{liu2024visual,
  title={Visual instruction tuning},
  author={Liu, Haotian and Li, Chunyuan and Wu, Qingyang and Lee, Yong Jae},
  journal={Advances in neural information processing systems},
  volume={36},
  year={2024}
}

@inproceedings{wu2023multimodal,
  title={{Multimodal large language models: A survey}},
  author={Wu, Jiayang and Gan, Wensheng and Chen, Zefeng and Wan, Shicheng and Philip, S Yu},
  booktitle={IEEE International Conference on Big Data },
  pages={2247--2256},
  year={2023},
  organization={IEEE}
}

@article{kumar2024performance,
  title={{Performance evaluation of SLAM-ASR: The Good, the Bad, the Ugly, and the Way Forward}},
  author={Kumar, Shashi and Thorbecke, Iuliia and Burdisso, Sergio and Villatoro-Tello, Esa{\'u} and Hacio{\u{g}}lu, Kadri and Rangappa, Pradeep and Motlicek, Petr and Ganapathiraju, Aravind and Stolcke, Andreas and others},
  journal={arXiv preprint arXiv:2411.03866},
  year={2024}
}

@article{cappellazzo2024largelanguagemodelsstrong,
      title={{Large Language Models Are Strong Audio-Visual Speech Recognition Learners}}, 
      author={Umberto Cappellazzo and Minsu Kim and Honglie Chen and Pingchuan Ma and Stavros Petridis and Daniele Falavigna and Alessio Brutti and Maja Pantic},
      year={2024},
      eprint={2409.12319},
      archivePrefix={arXiv},
      primaryClass={cs.CV},
      
}

@article{Martins2024EuroLLMML,
  title={EuroLLM: Multilingual Language Models for Europe},
  author={Pedro Henrique Martins and Patrick Fernandes and Joao Alves and Nuno M. Guerreiro and Ricardo Rei and Duarte M. Alves and Jos{\'e} P. Pombal and Amin Farajian and Manuel Faysse and Mateusz Klimaszewski and Pierre Colombo and Barry Haddow and Jos{\'e} Guilherme Camargo de Souza and Alexandra Birch and Andr{\'e} Martins},
  journal={ArXiv},
  year={2024},
  volume={abs/2409.16235},
  
}

@misc{gonzalezagirre2025salamandratechnicalreport,
      title={Salamandra Technical Report}, 
      author={Aitor Gonzalez-Agirre and Marc Pàmies and Joan Llop and Irene Baucells and Severino Da Dalt and Daniel Tamayo and José Javier Saiz and Ferran Espuña and Jaume Prats and Javier Aula-Blasco and Mario Mina and Adrián Rubio and Alexander Shvets and Anna Sallés and Iñaki Lacunza and Iñigo Pikabea and Jorge Palomar and Júlia Falcão and Lucía Tormo and Luis Vasquez-Reina and Montserrat Marimon and Valle Ruíz-Fernández and Marta Villegas},
      year={2025},
      eprint={2502.08489},
      archivePrefix={arXiv},
      primaryClass={cs.CL},
      
}

@article{Cheng2024ExploringTI,
  title={Exploring the Impact of Data Quantity on ASR in Extremely Low-resource Languages},
  author={Yao-Fei Cheng and Li-Wei Chen and Hung-Shin Lee and Hsin-Min Wang},
  journal={ArXiv},
  year={2024},
  volume={abs/2409.08872},
  
}

@article{bai2024seed,
  title={Seed-asr: Understanding diverse speech and contexts with llm-based speech recognition},
  author={Bai, Ye and Chen, Jingping and Chen, Jitong and Chen, Wei and Chen, Zhuo and Ding, Chuang and Dong, Linhao and Dong, Qianqian and Du, Yujiao and Gao, Kepan and others},
  journal={arXiv preprint arXiv:2407.04675},
  year={2024}
}

@inproceedings{fathullah2024prompting,
  title={Prompting large language models with speech recognition abilities},
  author={Fathullah, Yassir and Wu, Chunyang and Lakomkin, Egor and Jia, Junteng and Shangguan, Yuan and Li, Ke and Guo, Jinxi and Xiong, Wenhan and Mahadeokar, Jay and Kalinli, Ozlem and others},
  booktitle={ICASSP 2024-2024 IEEE International Conference on Acoustics, Speech and Signal Processing (ICASSP)},
  pages={13351--13355},
  year={2024},
  organization={IEEE}
}

@article{mittal2024salsa,
  title={Salsa: Speedy asr-llm synchronous aggregation},
  author={Mittal, Ashish and Prabhu, Darshan and Sarawagi, Sunita and Jyothi, Preethi},
  journal={arXiv preprint arXiv:2408.16542},
  year={2024}
}

\end{document}